\documentclass[12pt]{JHEP3}
\pdfoutput=1
\usepackage{graphicx}

\usepackage{epsfig}
\usepackage{amssymb}
\usepackage{bm}
\usepackage{amsmath}
\newcommand{\bea}{\begin{eqnarray}}
\newcommand{\eea}{\end{eqnarray}}

\makeatletter
\@addtoreset{equation}{section}
\makeatother

\title{
Impurities in Holography and Transport Coefficients \\
}
\author{
Koji Hashimoto$^{1}$ and 
Norihiro Iizuka$^{2}$
\\

$^1$
{\it Mathematical Physics Lab., RIKEN Nishina Center,
Saitama 351-0198, Japan}\\
E-mail: \email{koji(at)riken.jp}\\ 

$^2$
{\it Theory Division, CERN, CH-1211 Geneva 23, Switzerland}\\
E-mail: \email{norihiro.iizuka(at)cern.ch}\\
}

\abstract{
We present a way to include  impurities in AdS/CFT correspondence, 
in view of its application to condensed matter physics. 
Examples of these are the current impurity and spin impurity.  
We calculate electric conductivity and spin susceptibility 
of holographic superconductors, with doping of density/spin impurities.}

\preprint{
{\normalsize CERN-PH-TH-2012-198} \\
{\normalsize RIKEN-MP-51}
}

\begin{document}
\setcounter{page}{1}

\section{Introduction}
\label{sec1}

Holography or AdS/CFT correspondence \cite{Maldacena:1997re, Gubser:1998bc, Witten:1998qj} is an extremely useful tool, 
in the sense that it gives a new view of the hard-core problem of quantum field theory, 
strongly coupled limit of the theory, from the gravitational theory or string theory viewpoint. 
Recently there are many advances
applying the holography to more real-world setup like nuclear physics, hadron physics, and 
condensed matter physics such as quantum hall effects and superconductivity. 
For the case of condensed matter application, in many of the holographic setting, 
even though its validity to real-world condensed matter system is not clear, 
there is a big hope that these totally new holographic perspectives 
will shed new light on long-standing problems of these fields in the end. 
And in fact, recent progress along these directions, especially condensed matter physics, 
is quite remarkable. 
For reviews of recent development along these fields, see for example \cite{SachdevMueller,Hartnoll:2009sz,Herzog:2009xv,McGreevy:2009xe,Horowitz:2010gk,Sachdev:2010ch, Hartnoll:2011fn}. 

Many of the transport coefficients in strongly coupled systems can be calculated 
by using holographic setting. 
In condensed matter physics, there is a crucial ingredient, the effects of impurities.  
Adding impurities is very ubiquitous technique 
in condensed matter physics, it changes the basic nature of materials and 
changes the transport properties drastically. In other words, it changes the 
Hamiltonian of the system.  
In a very famous example of high $T_c$ cuprate superconductors, by compounding Sr or Ba
to La-Cu-O system to add the 
conductance charge carriers, the properties of the material  drastically changes from a
Mott insulator to a high $T_c$ superconductor.   
Another famous example is the Kondo effect, where magnetic impurities, {\it i.e.,} spins, 
make the resistance increase logarithmically as we lower the temperature. 
As is seen by these drastic changes of the properties of the material, 
impurities can induce many peculiar effects to the material. 
Therefore, it is very natural to ask, how the 
impurities give effects generically, 
in many of the holographic condensed matter settings which were studied before. 
In this paper we will take the first step for this direction, especially the effects of the 
 impurities, which include magnetic impurities. 

Kondo effect is induced due to the quantum spin nature of magnetic impurities coupled to 
conducting electrons near their Fermi-surfaces. 
The spin operator of the impurity fermion $\psi_{\rm im}$\footnote{This impurity fermion 
$\psi_{\rm im}$ should not be confused with conducting quasi-particle fermions, like electrons.} can be described by 
\bea
\langle \psi_{\rm im}^\dagger  \sigma^\mu \psi_{\rm im} \rangle \sim  J^\mu \,.
\eea
In this paper, treating $J^\mu$ as an  
 external
 input, we consider the effects of the impurity $J^\mu$ to the transport coefficients 
in holographic settings. Our impurity is introduced by hand. 
For simplicity, we consider only 
homogeneous impurities which induce homogeneous source current $J^\mu$.  
This naturally induces a bulk current $J^\mu(r)$, which is 
dependent only on the AdS bulk radial coordinates $r$. 

The organization of this paper is as follows; 
In section \ref{sec2}, 
we first describe in detail our method to include impurities in the AdS/CFT correspondence.
Then in section \ref{sec3}, 
we use the holographic superconductor model \cite{HHH} and introduce the impurity.
We calculate the AC conductivity, and evaluate the change of it due to the impurity.
We treat two kinds of impurities there: density impurity and spin impurity.
In section \ref{sec:spin}, we calculate the spin susceptibility in the same manner. 
In both cases, we clearly see the effect of the impurity. 
Section \ref{sec:disc} is for the discussion.


\section{Impurities in AdS/CFT}
\label{sec2}

Impurities play an indispensable role in condensed matter physics, 
while its treatment in the AdS/CFT approach has not been fully established. 
There are several interesting works \cite{Harrison:2011fs, Benincasa:2012wu} which introduced {\it local} impurities, but in general, impurities 
can be distributed all over the material of concern. Here we present our generic method to include some 
 impurities
in a general setup in AdS/CFT correspondence\footnote{See also \cite{Hartnoll:2008hs}, where the authors studied the momentum relaxation due to impurities.}.

\subsection{How to include the impurities}
\label{sec2-1}

Without the impurities, the conductivity (or the current correlator)  is treated in the AdS/CFT correspondence
as follows. Let us consider a strongly coupled fermion $\psi$ in $1+2$ dimensions. 
The fermion number current is given by ${\cal O}^\mu \equiv \bar{\psi} \gamma^\mu \psi$.  
We weakly gauge this $U(1)$ symmetry, which is nothing but the electromagnetism, to have the 
coupling $\int d^3x A_\mu {\cal O}^\mu$. To derive the current correlator, in the gravity side,
we upgrade the ``source" $A_\mu(x^\mu)$ for the operator ${\cal O}^\mu$ to a bulk field $A_M(x^\mu,r)$ where $r$ is the bulk
coordinate along the AdS radial direction. According to the AdS/CFT dictionary,
we demand that the bulk field approaches to the value of the boundary source,
$\lim_{r\rightarrow\infty }A_{M = \mu} = A_\mu $. We solve the equation in the bulk, and substitute it
to the bulk action, to obtain the classical bulk partition function which is a function of the
boundary source $A_\mu$. In the AdS/CFT \cite{Gubser:1998bc, Witten:1998qj}, this is equal to the boundary partition function
with the source $A_\mu$ of the source term $\int d^3x A_\mu {\cal O}^\mu$.

Now, let us add the impurities. 
Suppose the impurity is giving the electric density and the electric 
currents only, 
then the impurity coupling at the boundary theory is the minimal coupling 
$\int d^3x A_\mu {\cal O}_{\rm imp}^\mu$, where
the impurity operator would be given by the impurity fermion field $\psi_{\rm im} (x^\mu)$ as
${\cal O}_{\rm imp}^\mu \equiv \bar{\psi}_{\rm im}  \gamma^\mu \psi_{\rm im} $.  
Let us treat the impurities as classical objects. Since we can distribute the impurity 
in an arbitrary manner in real-world experiments, 
we can take the vacuum expectation value of the current 
$\langle {\cal O}_{\rm imp}^\mu (x^\mu)\rangle$
to be an arbitrary function. 
That is, our impurity coupling is 
\begin{eqnarray}
\int d^3x A_\mu(x) \langle {\cal O}_{\rm imp}^\mu(x^\mu)\rangle.
\label{boucoup}
\end{eqnarray}

In going to the gravity side of the AdS/CFT correspondence, 
we upgrade this coupling \eqref{boucoup} to the bulk coupling, 
\begin{eqnarray}
\int d^3xdr \;  A_M(x,r) J^M (x^\mu,r).
\label{sourcecoup}
\end{eqnarray}
We propose that this is a generic effect due to the impurities in AdS/CFT correspondence.
If one specifies the dynamics of the impurity, the configuration $J^M(x^\mu,r)$ is 
determined. Basically, the radial dependence of the bulk source field $J^M(x^\mu,r)$
represents how the impurity responds in different energy scales. Since 
the behavior of the impurities at different energy scale can be taken arbitrary as an 
external input, we take $J^M$ as an input source to the bulk gravitational action. 

To gain more insight on the reason of this bulk source coupling, 
let us consider the impurity in the following manner.
First, generic impurities are things put by hand. 
So we may allow such arbitrariness. Second, impurities can be heavy compounds 
and then their motion is not taken into account for transport coefficients. 
So the effect of the impurities are fixed in
the bulk, and affects directly the currents of the conducting electrons 
(which are gauge bosons in the bulk). 
This allows us to treat $J^M$ as a classical input. 
In section \ref{sec2-3}, we provide explicit models to derive the 
coupling \eqref{sourcecoup}, as an example.

In this way we consider the effect of the impurity
in the form of the function $J^M(x^\mu,r)$ which we give by hand.


\subsection{Spin impurities}

In most interesting cases, impurities couple to the electromagnetic fields (and to the conducting electrons and quasiparticles in strongly correlated systems) through a spin magnetic coupling. 
In the previous subsection we have treated a covariant and canonical coupling between the impurity fermion and the electromagnetic field. 

Let us consider the spin magnetic
interaction, which is written as 
\begin{eqnarray}
\int d^3x \; B_i(x) \langle {\cal S}^i (x^\mu) \rangle ,
\end{eqnarray}
where ${\cal S}^i(x^\mu)$ is the spin operator made from the impurity fermion field $\psi_{\rm im}$, and $B_i(x)$ is the magnetic 
field.\footnote{Here for simplicity the material is supposed to have  2 spatial dimensions, but one can generalize this to 3 spatial dimensions.
Note that even in 2 spatial dimensions, on materials one can introduce 3-dimensional spins.}
In terms of the field strength, this coupling is written as
\begin{eqnarray}
\int d^3x \; \frac12 \epsilon_{ijk} F_{jk}(x) \langle {\cal S}^i (x^\mu) \rangle 
=- \int d^3x \;  A_k (x) \langle \epsilon_{ijk} \partial_j {\cal S}^i (x^\mu) \rangle ,
\label{spinb}
\end{eqnarray}
which is of the form \eqref{boucoup}. So, introducing the spin magnetic coupling for the impurity is included in the 
scheme of the previous subsection.

In this paper, in order to show explicit examples, we consider a certain specific form of the impurity source $J^M(x^\mu,r)$ in the bulk (which we give by hand), namely, 
\begin{eqnarray}
J^M = c/r^n,
\label{jr}
\end{eqnarray}
and compute the AC conductivity in the $x$ direction. 
Obviously one can generalize the functional form of the impurity and
also the direction of the currents. 
For numerical calculations, we specifically treat the case of
$n=6$ as explicit examples. 
We have to require that the back reaction of the source $J$ may not change the asymptotic geometry,
so we choose the source $J$ which quickly decays near the boundary $r=\infty$. The radial $r$ dependence in the
geometry corresponds to the energy dependence of the impurity source.\footnote{If one regards the energy as a hypothetical 
temperature of the impurity, the asymptotic decay of the source $J$ means that near the 
boundary, where the energy (temperature) scale is large, 
the effect of the impurity is lower.}

According to \eqref{spinb}, the source $J^y$ is related to the spin  roughly as $J^y \sim \partial_x {\cal S}^z$.
So, if we define a bulk version of the source spin, it is (for our choice of $J$ as \eqref{jr}) ${\cal S}^z \sim  x/r^6$.
For numerical purpose, in this paper we consider only this case in section 3 and 4.


\subsection{Impurity models in holography}
\label{sec2-3}

In section \ref{sec2-1}, we have argued that the way we introduce the impurity 
is just the introduction of 
the source term in the bulk, \eqref{sourcecoup}. Here, for concreteness, we shall provide explicit
AdS/CFT models in which the impurity supplies the bulk source term.

Let us consider a theory in 1+2 dimensions, with a strongly correlated fermion $\psi$ and a strongly correlated
massive impurity fermion $\psi_{\rm im}$. According to the gauge/gravity correspondence, we consider operators
of our concern, 
\begin{eqnarray}
{\cal O}^\mu = \bar{\psi} \gamma^\mu \psi, \quad
{\cal O^\mu_{\rm im}} = \bar{\psi}_{\rm im} \gamma^\mu\psi_{\rm im}.
\end{eqnarray}
These are just fermion number current for each fermion.\footnote{For example, one can imagine
an $SU(N)$ gauge theory with two kinds of fermions $\psi$ and $\psi_{\rm im}$, both of which are in 
the fundamental representation of the $SU(N)$. Then the model looks a QCD with two flavors. If we suppose
that the fermion $\psi$ is light while the impurity fermion $\psi_{\rm im}$ is superheavy, then
the model suffices our purpose. In this paper, we would like to consider more generic situation in 
holography.}
 The AdS/CFT correspondence requires 
bulk fields which corresponds to these fields,
\begin{eqnarray}
{\cal O}^\mu  \leftrightarrow A_\mu, \quad
{\cal O^\mu_{\rm im}} \leftrightarrow B_\mu.
\end{eqnarray}
Now, because these two currents are conserved, we write a gauge-invariant action for these two $U(1)$
gauge fields:
\begin{eqnarray}
L = -\frac14 F_{\mu\nu} F^{\mu\nu} -\frac14 H_{\mu\nu}H^{\mu\nu} 
+ cH_{\mu\nu}F^{\mu\nu},
\label{FH}
\end{eqnarray}
where
\begin{eqnarray}
F_{\mu\nu} = \partial_\mu A_\nu - \partial_\nu A_\mu, \quad
H_{\mu\nu} = \partial_\mu B_\nu - \partial_\nu B_\mu .
\end{eqnarray}
The last term in the Lagrangian is one example of a coupling between the conducting electron and the
impurity fermion.\footnote{The coupling is similar to the one for realizing a vector meson dominance in
hidden local symmetry models. We would like to thank K.~Fujikawa for pointing this out to us.} 
In general, we can allow arbitrary couplings which are not prohibited by the
symmetries, and this is the lowest order term in the derivative expansion.

If impurities are heavy, we can generally fix the dynamics of the impurity-induced gauge potential $B_\mu$.
So we can suppose that we may consider a background configuration for the bulk field $B_\mu$.
Once we allow it, substituting this $B_\mu$ configuration back into the action \eqref{FH}
and make an integration by parts, then we arrive at the bulk source coupling \eqref{sourcecoup}, $A_\mu J^\mu$ in the
bulk.\footnote{Strictly speaking, the electromagnetic $U(1)$ is a linear combination of the first and the second $U(1)$.
However, since the impurity dynamics can be killed due to its heaviness, it is equivalent to consider only the $A_\mu$ as dynamical field to calculate the
conductivity.}

The configuration of the source $J$ in the bulk depends on the dynamics of the impurity. For example, the asymptotic behavior
of the source $J$ would be related to the form of the coupling and also the conformal dimension of the impurity operator.
Furthermore, we may have a variety of the form of the couplings and also we may allow multiple kinds of the impurity operators
with different conformal dimensions. This means that basically an arbitrary configuration is allowed for the bulk source $J$,
depending on the models. Therefore in this paper we do not stack to a specific model, and we allow generic configuration of
the bulk source $J$. The idea is along the standard picture that one can control the distribution of the impurity source
by hand.

We present one more model which would suffice for introducing a spin background
according to the idea written above. The impurity spin operator is
\begin{eqnarray}
{\cal S}^i_{\rm im} = \bar{\psi}_{\rm im} \gamma^i\gamma^5\psi_{\rm im},
\label{spindensity}
\end{eqnarray}
in 1+3 dimensional notation for fermions. This can be easily seen if one 
decomposes the four-component spinor
$\psi_{\rm im}$ into 2+2 components $(\varphi,\chi)^T$, and explicitly write the gamma matrices 
in terms of the Pauli matrices:
${\cal S}^i = \varphi^\dagger \sigma^i \varphi + \chi^\dagger \sigma^i \chi$. So we can consider 
the spatial component of 
the axial current as a spin density. Let us introduce a bulk vector field ${B}_\mu$ which is 
dual to the operator 
${\cal S}^i$, in the same manner. Then, since the axial current is not 
conserved for the massive fermions,  
the corresponding $U(1)$ gauge symmetry for ${B}_\mu$  is broken in the bulk. 
Therefore we can write a model like
\begin{eqnarray}
L = -\frac14 F_{\mu\nu} F^{\mu\nu} -\frac14 H_{\mu\nu}H^{\mu\nu} 
+ \frac12 m^2 B_\mu B^\mu + cH_{\mu\nu}F^{\mu\nu}.
\label{FHmodel}
\end{eqnarray}
Note that we have the mass term which breaks the $U(1)$ gauge invariance for the field $B_\mu$,
so  $B_\mu$ is a massive gauge field (Proca field). The model, with the coupling $c$, explicitly introduces
the interaction between the conducting electrons and the impurity spins. The mass term can be replaced by
a condensation of a charged scalar field in the bulk, as in the case of the holographic superconductors \cite{HHH}.

The spin is a pseudo-vector, as the current \eqref{spindensity} is a pseudo-vector.
In the model \eqref{FHmodel}, the last term breaks the parity invariance. To avoid
such a violation of the parity symmetry, one can write other kind of the coupling.
For example, the following higher derivative coupling
\begin{eqnarray}
c \; \epsilon^{\mu\nu\rho\sigma} F_{\mu\nu} H_{\rho\sigma} B_\lambda B^\lambda
\end{eqnarray}
preserves the parity symmetry. (Note that without the last factor $B_\lambda B^\lambda$
the coupling is trivially zero through an integration by parts.)

It is of course possible to proceed with these explicit models more, 
but it is not our purpose of this paper: we argue that in this way
generic effects of the impurity can be taken into account as a bulk source term.
In the rest of this paper, we just take a single configuration of the source $J$ which is \eqref{jr} with $n=6$, as an example.


\section{Conductivity}
\label{sec3}

In this section, we calculate the conductivity of a superconductor, in the 
presence of the impurity. 
As we studied in the previous section,
the impurity can be introduced via the source term for the bulk gauge field. 
Once we include the source term,
the equations of motion for the bulk fields are modified.

In general, the introduction of the source term  changes not only the 
gauge field configuration but also the background geometry. 
It is desirable to calculate the back-reaction to the geometry,
but in this paper we shall not consider it, by taking the gravitational 
coupling constant to zero. In other words, we take a probe limit,
as the gauge sector works as a probe for the geometry.
The limit is useful in two senses: first, it can pick up intrinsic physics
of the electromagnetic currents as they live in the gauge field sector, 
and second, it offers a simpler ground for calculations. 

First, we solve the background configuration of the gauge fields and the scalar fields,
with the source term which represents the impurity. Then, we study
the fluctuation to calculate the conductivity. Our calculation follows what has been originally proposed in \cite{HHH} for the holographic superconductors.


\subsection{General strategy}

\subsubsection{The background configuration in the bulk}

The Lagrangian in the bulk is that of a Maxwell field and a charged scalar field,
\begin{eqnarray}
S = 
\int \! d^4x \sqrt{-g}
\left[
-\frac14 F_{\mu\nu}F^{\mu\nu} -|\partial \Psi -iA \Psi|^2 + \frac{2}{L^2}|\Psi|^2 
+A_\mu J^\mu\right] \, .
\label{actionJ}
\end{eqnarray}
Note that we have added the last term, the source term, which represents the
impurity.
The scalar potential $(2/L^2) |\Psi|^2$
can be of a different form, but we just follow the popular example 
presented in \cite{HHH}. 

The background metric is the Schwarzschild black brane,\footnote{
One may wonder if one can take an AdS-Reissner-N\"odrstrom black brane 
as a background metric, as anyway we are going to consider a nontrivial 
configuration of $A_0(r)$. However,   
with the Reissner-N\"odrstrom background, in the probe limit $\kappa \rightarrow 0$, 
the temporal component of the gauge field should go to infinity. Then it is 
difficult to make sense of the the fluctuation of the gauge field around such a
divergent background.
}
\begin{eqnarray}
&& ds^2 = -f(r) dt^2 + \frac{1}{f(r)} dr^2 + r^2 (dx^2+dy^2), 
\\
&&
f(r) = \frac{r^2}{L^2} - \frac{M}{r} \, .
\end{eqnarray}
The location of the horizon $r=r_+$ and the temperature $T$ of the black brane 
is
\begin{eqnarray}
r_+ = (L^2 M)^{1/3}, \quad 
T = \frac{3 M^{1/3}}{4\pi L^{4/3}}.
\end{eqnarray}
In this background metric, we solve the equation of motion of
\eqref{actionJ}, with a given source $J(r)$.

The equation of motion for the gauge field and the scalar field is
\begin{eqnarray}
\label{EOMtosolveone}
&& - \frac{1}{\sqrt{-g}}\partial_\mu (\sqrt{-g} F^{\mu\nu}) +  i g^{\nu \lambda} \left( \Psi^* (\partial_\lambda - i A_\lambda) \Psi - \Psi ( \partial_\lambda +  i A_\lambda) \Psi^*\right)  = J^\nu \,,
\\
&&
\label{EOMtosolvetwo}
\frac{1}{\sqrt{-g}}(\partial_\mu -iA_\mu )
\left[
\sqrt{-g} g^{\mu\nu} 
(\partial_\nu - i A_\nu) \Psi
\right]
 + \frac{2}{L^2} \Psi = 0\, .
\end{eqnarray}
We need to solve these equations to obtain the background configuration
of the gauge field and the scalar field. 

\subsubsection{Calculation of the conductivity}
Given the background solutions of equations (\ref{EOMtosolveone}) and (\ref{EOMtosolvetwo}), we can 
now calculate the conductivity.  
For the calculation of the conductivity, say, along the $x$ direction, 
we turn on a fluctuation of $A_x$ which is time-dependent with 
frequency $\omega$ as $\sim e^{- i \omega t}$. 
The equation for this $A_x$ fluctuation is
\begin{eqnarray}
A_x'' + \frac{f'}{f} A_x'  + \left(\frac{\omega^2}{f^2}-\frac{2\Psi^2}{f}
\right) A_x = 0  \, .
\label{fluceq}
\end{eqnarray}
Note that this is the same equation as what is given in \cite{HHH}, since
the source term which we introduce affects only the background
configuration.\footnote{If one introduces a source term which
is not linear in $A_x$ but quadratic in $A_x$, it will affect directly
to the fluctuation equation. The choice of the form of the term with the impurity
is dependent on what kind of interaction one wants to introduce in the 
boundary theory.}
The boundary condition to solve the fluctuation equation is
the in-going boundary condition at the horizon.

As explained in \cite{HHH}, the conductivity $\sigma(\omega)$ is
calculated by the ratio of the asymptotic coefficients of the 
normalizable and the non-normalizable modes of $A_x$. 
Expanding 
\begin{eqnarray}
A_x = A_x^{(0)} + \frac{A_x^{(1)}}{r} + \cdots \, \quad (r \sim \infty) \, 
\end{eqnarray}
the conductivity is given by
\begin{eqnarray}
\sigma(\omega) = - \frac{iA_x^{(1)}}{\omega A_x^{(0)}}.
\end{eqnarray}

In the following, we treat two examples: (i) the source is $J^t$, meaning that
the impurity affects the electric density, and (ii) the source is $J^y$, meaning
a particular kind of the impurity spin density.



\subsection{The case of impurity density}

First, we consider the example with $J^t(r)$. 
Obviously, for the background configuration of the gauge field, we need 
only the temporal component,
\begin{eqnarray}
A_t= \Phi(r).
\end{eqnarray}
We will take a gauge where $\Psi$ takes the real value. 

The equations of motion for $\Phi(r)$ and $\Psi(r)$ are easily derived as
\begin{eqnarray}
&&
\Phi'' + \frac{2}{r} \Phi' -\frac{2\Psi^2}{f} \Phi = J^t  \,, 
\\
&&
\Psi'' + \left(
\frac{f'}{f} + \frac{2}{r}\right) \Psi' + \frac{\Phi^2}{f^2} \Psi
+\frac{2}{L^2f} \Psi = 0 \,.
\end{eqnarray}
The boundary conditions we need to care about are
$\Phi(r=r_+)=0$ at the horizon, and at the asymptotic boundary $r \sim \infty$,
\begin{eqnarray}
&&
\Phi (r) \sim \mu - \frac{\rho}{r} + \cdots \, ,\\
&&
\Psi (r) \sim \frac{\Psi^{(1)}}{r} + \frac{\Psi^{(2)}}{r^2} + \cdots \, .
\end{eqnarray}
As chosen in \cite{HHH}, we put either $\Psi^{(1)}$ or $\Psi^{(2)}$ to vanish.
For simplicity in this paper, we put $\Psi^{(2)}=0$. Then the input variable is
only the charge-career density 
$\rho$ (and the temperature $T$ of the background geometry).
The chemical potential $\mu$ is determined by $\rho$.

\begin{figure}
\centering
\includegraphics[scale=0.7]{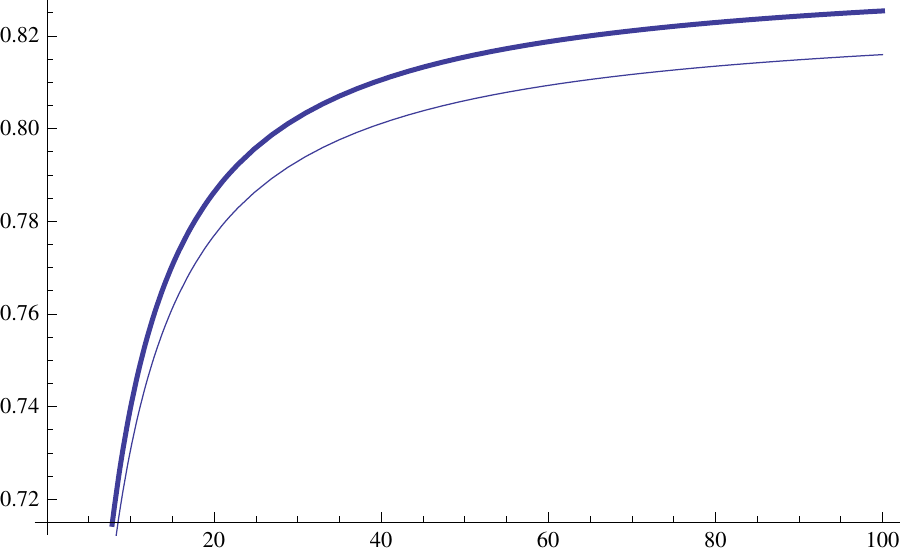} 
\put(-20,-5){$r$}
\put(-205,100){$\Phi(r)$}
\caption{The plot of the background configuration of $\Phi(r)$.
Thick line is with the impurity $J^t=0.1/r^6$, and the thin line
is without the impurity. }
\label{figPhi}
\end{figure}

For a given form of the source $J^t$, it is easy to perform a numerical
calculation for $\Phi(r)$ and $\Psi(r)$. The plot of the resultant configuration,
with and without the source term, is shown in Fig.~\ref{figPhi}.
For the numerical parameters, we have chosen $J^t = 0.1/r^6$, and 
$\rho=1$, $L=1$, and $M=0.2$. The temperature is $T \sim 0.14$ which is 
in the superconducting phase \cite{HHH}.

\begin{figure}
\centering
\includegraphics[scale=0.7]{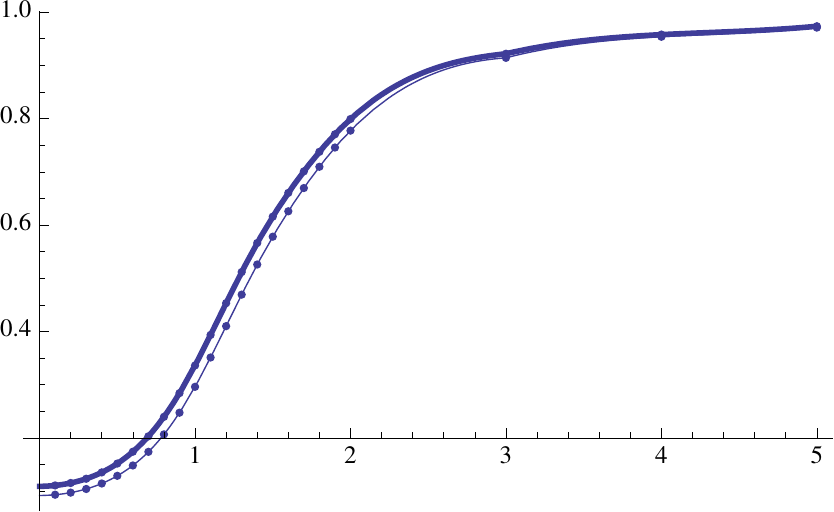} \hspace{10mm}
\includegraphics[scale=0.7]{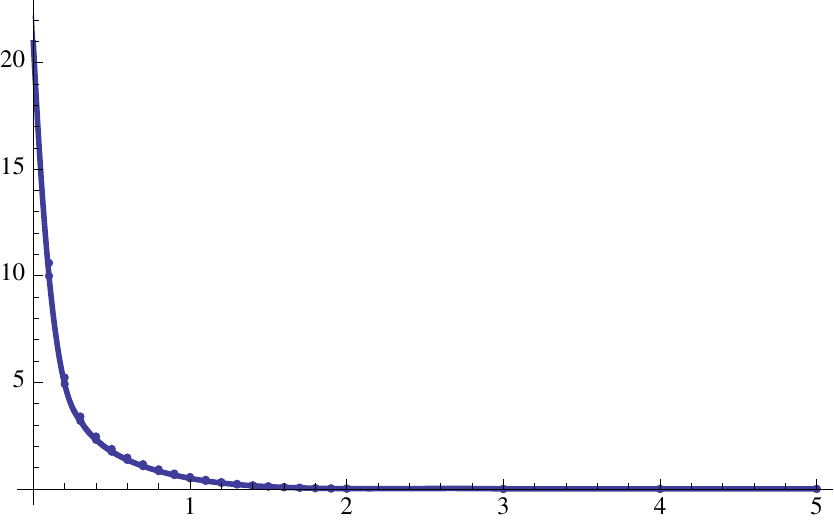}
\put(-220,-0){$\omega$}
\put(-20,-5){$\omega$}
\put(-400,100){Re[$\sigma$]}
\put(-195,100){Im[$\sigma$]}
\caption{The plot of the conductivity $\sigma$ as a function of the frequency $\omega$.
Thick lines: conductivity with the impurity $J^t=0.1/r^6$. 
Thin lines with dots: conductivity with no impurity. 
Left: Real part of $\sigma$. Right: Imaginary part of $\sigma$. }
\label{fig1}
\end{figure}

Next, we turn to solve the fluctuation equation \eqref{fluceq} to obtain the 
conductivity. 
It is straightforward to solve the fluctuation equation numerically,
and the result for the real and the imaginary parts of the conductivity is
shown in Fig.~\ref{fig1}.

Interestingly, the real part of the conductivity increases due to the impurity.
A possible interpretation of this effect is as follows: since we have introduced 
the impurity as a background density $J^t$, we may think of it as an increase of
an ``effective mass" of the background geometry. Once the mass of the black brane 
increases, the effective temperature increases. This generally increases the
conductivity, which is consistent with our numerical result.


\subsection{The case of impurity spin density}

We provide another example, which is $J^y(r)$. As explained earlier,
turning on the $J^y$ component represents a background spin configuration
caused by the impurity.

With this source term in the bulk, the $A_y$ component is excited 
as a background. The equations of motion are
\begin{eqnarray}
&&
\Phi'' + \frac{2}{r} \Phi' -\frac{2\Psi^2}{f} \Phi = 0  \,,
\\
\label{AyEOM}
&&
f A_y'' + f' A_y' - 2 \Psi^2 A_y + r^2 J^y = 0 \,,
\\
&&
\Psi'' + \left(
\frac{f'}{f} + \frac{2}{r}\right) \Psi' + \frac{\Phi^2}{f^2} \Psi - \frac{A_y^2}{r^2 f} \Psi
+\frac{2}{L^2f} \Psi = 0 \,.
\end{eqnarray}
The boundary condition for the $A_y$ is 
$A_y \sim {\cal O}(1/r)$ at $r\sim \infty$, as the background source
$J^y$ should induce a normalizable mode of $A_y$.

\begin{figure}
\centering
\includegraphics[scale=0.7]{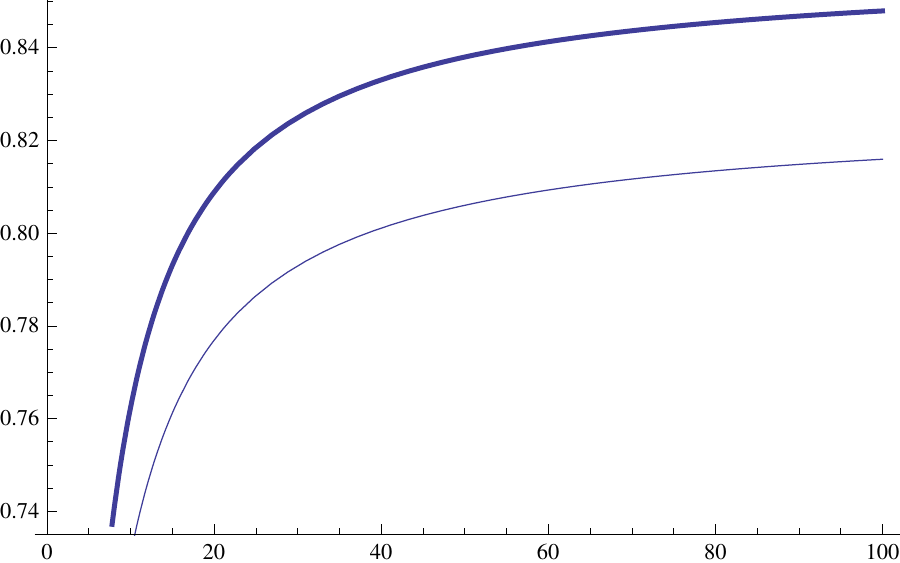} \hspace{10mm}
\includegraphics[scale=0.7]{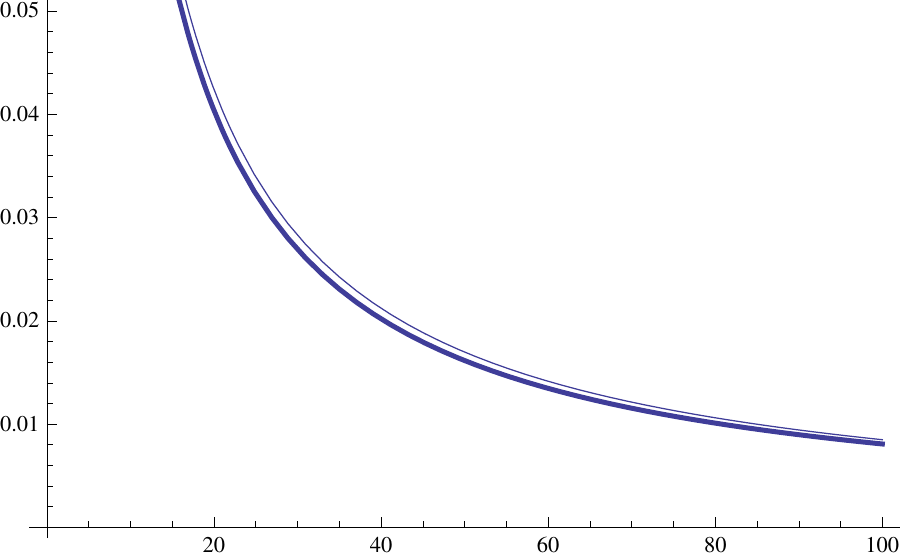}
\put(-240,-5){$r$}
\put(-20,-5){$r$}
\put(-420,100){$\Phi(r)$}
\put(-205,100){$\Psi(r)$}
\caption{The plot of the gauge configuration ($\Phi$, Left) and the scalar configuration ($\Psi$, Right).
Thick lines are with the impurity while the thin lines are without the impurity. 
 }
\label{fig2}
\end{figure}

\begin{figure}
\centering
\includegraphics[scale=0.7]{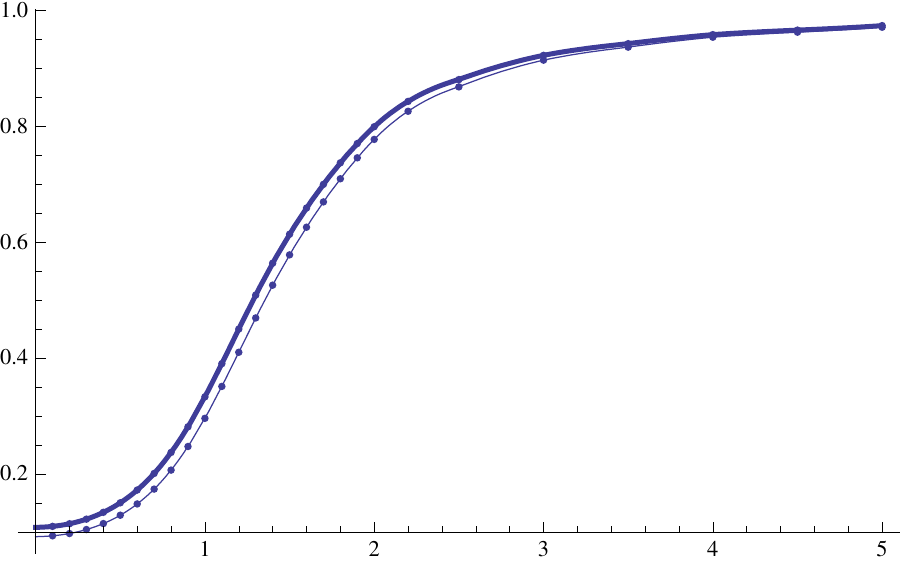} \hspace{10mm}
\includegraphics[scale=0.7]{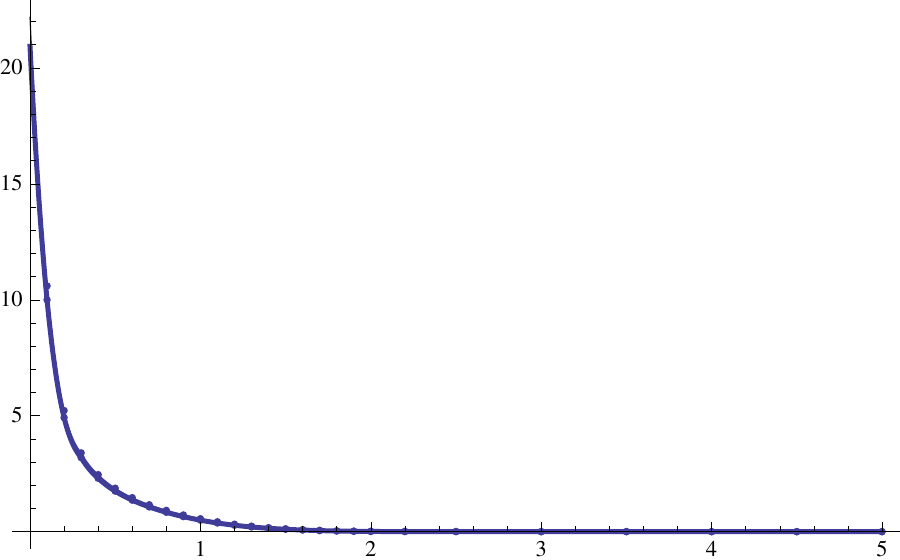}
\put(-240,-5){$\omega$}
\put(-20,-5){$\omega$}
\put(-425,100){Re[$\sigma$]}
\put(-205,100){Im[$\sigma$]}
\caption{The plot of the conductivity $\sigma$ as a function of the frequency $\omega$, for the
spin impurity $J^y(r)$.
Thick lines: conductivity with the impurity $J^y=0.1/r^6$. 
Thin lines with dots: conductivity with no impurity. 
Left: Real part of $\sigma$. Right: Imaginary part of $\sigma$. }
\label{fig2-2}
\end{figure}

The numerical calculation of the background gauge and scalar configuration, 
and also of the fluctuation \eqref{fluceq}, is straightforward, and the results are 
shown in Fig.~\ref{fig2} and Fig.~\ref{fig2-2}. Interestingly, the result looks quite similar
to what has been obtained for the case of $J^t$. The interpretation 
would be the same as that for $J^t$: the background impurity may
cause an effect of an increase of the ``effective temperature".  

Note that if we measure the conductivity along the $y$ direction,
the result is different. This is because the fluctuation 
of $A_y$ couples $\delta\Psi$, as the background $A_y$ is nonzero.
In addition, 
once we work without the probe limit, there should appear a difference between
the conductivity along the $x$ and the $y$ directions.

\section{Spin susceptibility}
\label{sec:spin}

In this section, we calculate the effect of the impurity on the spin susceptibility.  
As we have described in section \ref{sec2}, 
the bulk source $J^y$ we have introduced may be considered as
a spin impurity. So it should have some effect on the spin susceptibility in the measurement.

The spin susceptibility can be measured by a correlator of magnetic fields. Its treatment in the 
context of the AdS/CFT correspondence was discussed in, for example, \cite{Luo:2012am}. 
The magnetic field fluctuation is a kind of the gauge field fluctuation which we considered in the previous 
section for computing the AC conductivity. The only difference is the fact that we need a spatial
modulation for $A_\mu$ to induce the typical magnitude of the fluctuation magnetic field: Concretely, we consider
the fluctuation of the form
\begin{eqnarray}
\delta A_x = e^{-i\omega t} e^{iqy}a(r),
\end{eqnarray}
where $q$ is the momentum along the $y$ direction. This is equivalent to consider the fluctuation of the
magnetic field,\footnote{
We may consider a different orientation for the spatial modulation. For example, even for the same spin direction
$B_3$, we may take
$\delta A_y = e^{-i\omega t} e^{iqx}a(r)$, which gives the magnetic field fluctuation 
$\delta B_3 = \partial_x \delta A_y = i q e^{-i\omega t} e^{iqx}a(r)$.
}
\begin{eqnarray}
\delta B_3 = -\partial_y \delta A_x = -i q e^{-i\omega t} e^{iqy}a(r).
\label{spinfl}
\end{eqnarray}
The ratio of the normalizable and the non-normalizable modes gives the spin susceptibility, so, the
procedure to obtain the spin susceptibility is the same as that for the conductivity.

\begin{figure}
\centering
\includegraphics[scale=0.7]{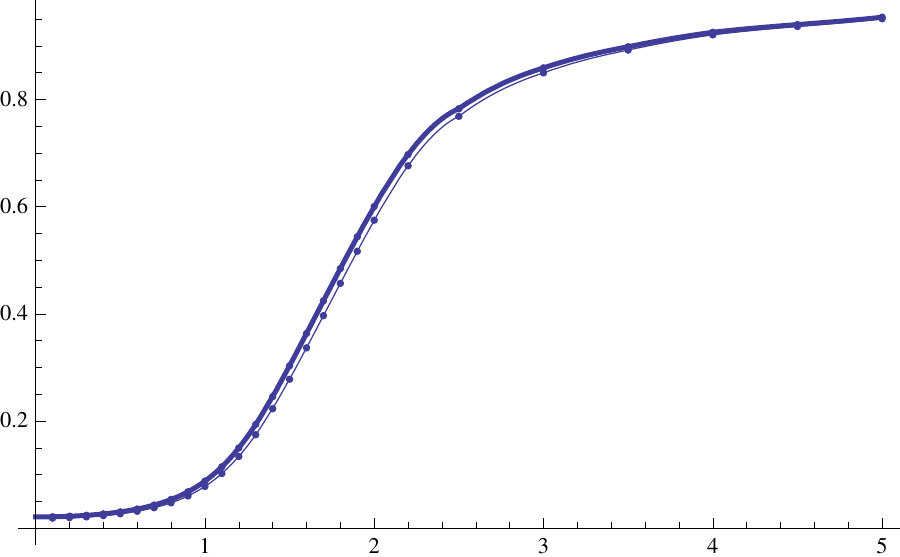} \hspace{10mm}
\includegraphics[scale=0.7]{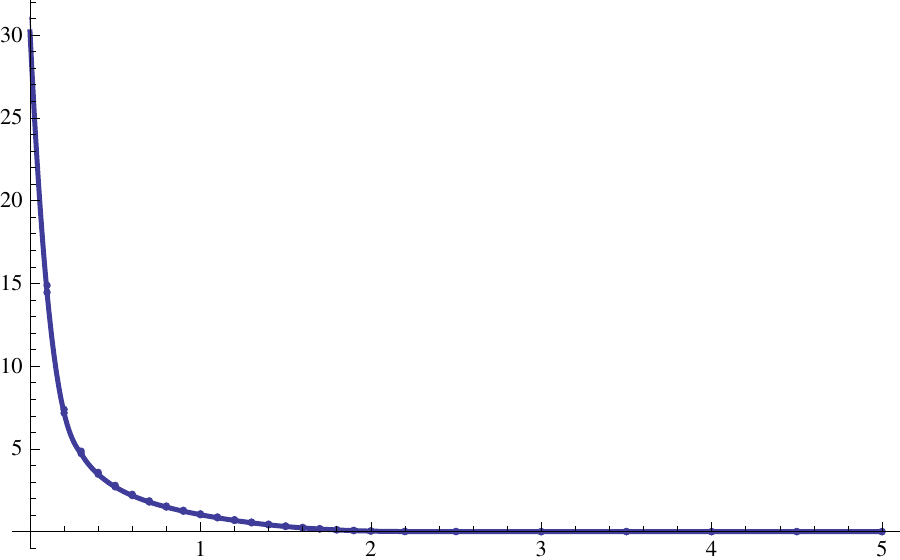}
\put(-220,-5){$\omega$}
\put(-20,-5){$\omega$}
\put(-380,100){Re[$\chi_s$]}
\put(-165,100){Im[$\chi_s$]}
\caption{Rigid thick line is the spin susceptibility with the impurity, while the thin 
line is for no impurity.
We have chosen the impurity $J^y=0.2/r^6$ as before, and took the spatial modulation 
$q=1$ for simplicity. 
The real part of the spin susceptibility increases for the middle region of the frequency $\omega$
under the doping of the impurity. }
\label{figspin1}
\end{figure}

The fluctuation should satisfy the on-shell equation. Substituting the fluctuation \eqref{spinfl} and the background geometry to the on-shell equation, we obtain the following equation
\begin{eqnarray}
\frac{1}{\sqrt{-g} g^{tt}g^{xx}} \partial_r \left(
\sqrt{-g} g^{rr} g^{xx} \partial_r a(r)\right)
- q^2 \frac{g^{yy}}{g^{tt}}a(r)   - 2 \frac{\Psi^2}{g^{tt}} a(r)= \omega^2 a(r) \,.
\end{eqnarray}
The numerical calculation of the spin susceptibility is straightforward.
The result is shown in Fig.\ref{figspin1}. We observe that the spin susceptibility $\chi_s$ increases due to the
impurity doping.

\section{Discussions}
\label{sec:disc}

In this paper, we proposed a generic method to include 
impurities in 
the context of AdS/CFT in application to condensed matter physics. 
We have calculated several examples for the 
transport coefficients in a  superconducting phase: AC conductivities and spin susceptibility. 
The effects are clearly seen.

The essence is to introduce a source term in the bulk side of the AdS/CFT correspondence
which sources the bulk gauge field. The bulk source term may have an 
arbitrary profile,
which corresponds to the arbitrariness of how we introduce the impurities to the system.
In section \ref{sec2}, we have given explicit models which can determine the 
profile of the bulk source term from the property of the impurity, but they are
just examples: in principle, we may allow arbitrary profile for the bulk source term
in general.

For the example which we analyzed in this paper, the impurity in the holographic superconductors, we observed that the
AC conductivity and the spin susceptibility increase due to the impurity. 
The former increases at the lower frequency, while the latter
seems to be sensitive to the typical spatial momentum of the magnetic field.
Since the impurity is introduced as a small perturbation, it is natural
that the impurity affects only the low frequency region of the AC conductivity. 
The increase may be interpreted as an increase of an ``effective temperature" of
the system. We have done the calculation with different bulk
profile for the bulk source field, and obtained a similar behavior. 
It could be 
that the increase at the low frequency region which we obtained is a generic
feature, but we do not have a clear understanding for this. 

We have worked in a probe limit where the effects of the 
gravity 
is ignored. 
In that case, the background geometry unperturbed with the impurity source term
would be a Reissner-N\"odrstrom black brane with a scalar hair \cite{Hartnoll:2008kx}. 
There are two effects by the gravity effects. One is the back-reaction to the background geometry 
due to the impurity. The other is the gravity fluctuation on the background, which 
couples to the gauge field fluctuation for the conductivity calculation.

One can also introduce other fields, like dilaton and axion, in the bulk action 
to simulate quantum Hall systems as in \cite{Goldstein:2010aw,Bayntun:2010nx}.
If we have running dilaton, the geometry can be replaced by Lifshitz type of geometry 
or others \cite{Kachru:2008yh, Goldstein:2009cv, Charmousis:2010zz, Iizuka:2011hg} 
at IR where the geometry shows different symmetries, 
such as non-relativistic one. In this way, on the gravity side,
we have varieties of generalizations. 
In the holographic superconductors, some impurity of course would change the
critical temperatures, which would be interesting, and that could be calculated when
one includes the coupling to the gravity. Various transport coefficients are
affected by the impurities, while the universality of the critical exponents against
perturbations by the impurities would be an interesting question.

The impurity we considered in this paper is ``classical" impurities, while in reality
impurity may be caused by quantum impurity fields. If one allow quantum impurities,
nontrivial dynamics of the impurities would determine the total effect to the conducting
quasiparticles. For example, one-loop effect of the impurity fermions would become 
the leading order effect \cite{Sachdev:2011ze, Hashimoto:2012ti}.

The treatment of the bulk source field in this paper is in the Maxwell sector, so the
effect of the impurity looks just a linear order in $J$. 
However, the coupling to the scalar field is nonlinear. Or, once
one couples it to a dynamical gravity, higher order term in $J$ will show up,
due to the nonlinearity of the gravity. 
There are interesting effects 
which appear only at higher order in the impurity perturbation, such as the renowned
Kondo effect. 
As we pointed out at the end of section 3, there should be anisotropy effects for the conductivity 
due to the non-linearity. 
It is interesting to investigate these furthermore. 
The exact treatment of the bulk source field in the dynamical gravity model
is intimately related to the
dynamics of the impurity, so solving models beyond the probe limit 
is of importance.


\acknowledgments
We would like to thank Akira Furusaki, Nilay Kundu, Prithvi Narayan, Shiroman Prakash, and Sandip P. Trivedi  for valuable helpful conversation. 
K.H.~is partly supported by
the Japan Ministry of Education, Culture, Sports, Science and
Technology. 
N.I. would like to thank Mathematical physics laboratory in RIKEN for very kind hospitality. 
N.I. is supported in part by the COFUND fellowship at CERN. 



\begin{thebibliography}{10}

\bibitem{Maldacena:1997re} 
  J.~M.~Maldacena,
  ``The Large N limit of superconformal field theories and supergravity,''
  Adv.\ Theor.\ Math.\ Phys.\  {\bf 2}, 231 (1998)
  [hep-th/9711200].

\bibitem{Gubser:1998bc} 
  S.~S.~Gubser, I.~R.~Klebanov and A.~M.~Polyakov,
  ``Gauge theory correlators from noncritical string theory,''
  Phys.\ Lett.\ B {\bf 428}, 105 (1998)
  [hep-th/9802109].
  
\bibitem{Witten:1998qj} 
  E.~Witten,
  ``Anti-de Sitter space and holography,''
  Adv.\ Theor.\ Math.\ Phys.\  {\bf 2}, 253 (1998)
  [hep-th/9802150].


  
\bibitem{SachdevMueller}
S.~Sachdev and M.~Mueller, 
``Quantum criticality and black holes,'' 
arXiv:0810.3005 [cond-mat.str-el].



\bibitem{Hartnoll:2009sz} 
  S.~A.~Hartnoll,
  ``Lectures on holographic methods for condensed matter physics,''
  Class.\ Quant.\ Grav.\  {\bf 26}, 224002 (2009)
  [arXiv:0903.3246 [hep-th]].
  
\bibitem{Herzog:2009xv} 
  C.~P.~Herzog,
  ``Lectures on Holographic Superfluidity and Superconductivity,''
  J.\ Phys.\ A A {\bf 42}, 343001 (2009)
  [arXiv:0904.1975 [hep-th]].

\bibitem{McGreevy:2009xe} 
  J.~McGreevy,
  ``Holographic duality with a view toward many-body physics,''
  Adv.\ High Energy Phys.\  {\bf 2010}, 723105 (2010)
  [arXiv:0909.0518 [hep-th]].


\bibitem{Horowitz:2010gk} 
  G.~T.~Horowitz,
  ``Introduction to Holographic Superconductors,''
  arXiv:1002.1722 [hep-th].

\bibitem{Sachdev:2010ch} 
  S.~Sachdev,
  ``Condensed Matter and AdS/CFT,''
  arXiv:1002.2947 [hep-th].


\bibitem{Hartnoll:2011fn} 
  S.~A.~Hartnoll,
  ``Horizons, holography and condensed matter,''
  arXiv:1106.4324 [hep-th].



  
\bibitem{HHH}
  S.~A.~Hartnoll, C.~P.~Herzog and G.~T.~Horowitz,
  ``Building a Holographic Superconductor,''
  Phys.\ Rev.\ Lett.\  {\bf 101}, 031601 (2008)
  [arXiv:0803.3295 [hep-th]].
  

\bibitem{Harrison:2011fs} 
  S.~Harrison, S.~Kachru and G.~Torroba,
  ``A maximally supersymmetric Kondo model,''
  arXiv:1110.5325 [hep-th].
  
\bibitem{Benincasa:2012wu} 
  P.~Benincasa and A.~V.~Ramallo,
  ``Holographic Kondo Model in Various Dimensions,''
  arXiv:1204.6290 [hep-th].
  
\bibitem{Hartnoll:2008hs} 
  S.~A.~Hartnoll and C.~P.~Herzog,
  ``Impure AdS/CFT correspondence,''
  Phys.\ Rev.\ D {\bf 77}, 106009 (2008)
  [arXiv:0801.1693 [hep-th]].
  
\bibitem{Luo:2012am} 
  M.~J.~Luo,
  ``Dynamic Scaling of Holographic Spin Fluctuations,''
  arXiv:1205.3267 [hep-th].


\bibitem{Hartnoll:2008kx} 
  S.~A.~Hartnoll, C.~P.~Herzog and G.~T.~Horowitz,
  ``Holographic Superconductors,''
  JHEP {\bf 0812}, 015 (2008)
  [arXiv:0810.1563 [hep-th]].
  
\bibitem{Goldstein:2010aw} 
  K.~Goldstein, N.~Iizuka, S.~Kachru, S.~Prakash, S.~P.~Trivedi and A.~Westphal,
  ``Holography of Dyonic Dilaton Black Branes,''
  JHEP {\bf 1010}, 027 (2010)
  [arXiv:1007.2490 [hep-th]].
  
\bibitem{Bayntun:2010nx} 
  A.~Bayntun, C.~P.~Burgess, B.~P.~Dolan and S.~-S.~Lee,
  ``AdS/QHE: Towards a Holographic Description of Quantum Hall Experiments,''
  New J.\ Phys.\  {\bf 13}, 035012 (2011)
  [arXiv:1008.1917 [hep-th]].

\bibitem{Kachru:2008yh} 
  S.~Kachru, X.~Liu and M.~Mulligan,
  ``Gravity Duals of Lifshitz-like Fixed Points,''
  Phys.\ Rev.\ D {\bf 78}, 106005 (2008)
  [arXiv:0808.1725 [hep-th]].

  
\bibitem{Goldstein:2009cv} 
  K.~Goldstein, S.~Kachru, S.~Prakash and S.~P.~Trivedi,
  ``Holography of Charged Dilaton Black Holes,''
  JHEP {\bf 1008}, 078 (2010)
  [arXiv:0911.3586 [hep-th]].
  
\bibitem{Charmousis:2010zz} 
  C.~Charmousis, B.~Gouteraux, B.~S.~Kim, E.~Kiritsis and R.~Meyer,
  ``Effective Holographic Theories for low-temperature condensed matter systems,''
  JHEP {\bf 1011}, 151 (2010)
  [arXiv:1005.4690 [hep-th]].
  
\bibitem{Iizuka:2011hg} 
  N.~Iizuka, N.~Kundu, P.~Narayan and S.~P.~Trivedi,
  ``Holographic Fermi and Non-Fermi Liquids with Transitions in Dilaton Gravity,''
  JHEP {\bf 1201}, 094 (2012)
  [arXiv:1105.1162 [hep-th]].
  
  
\bibitem{Sachdev:2011ze} 
  S.~Sachdev,
  ``A model of a Fermi liquid using gauge-gravity duality,''
  Phys.\ Rev.\ D {\bf 84}, 066009 (2011)
  [arXiv:1107.5321 [hep-th]].
  
\bibitem{Hashimoto:2012ti} 
  K.~Hashimoto and N.~Iizuka,
  ``A Comment on Holographic Luttinger Theorem,''
  arXiv:1203.5388 [hep-th].
  
    
\end{thebibliography}
\end{document}